\documentclass[runningheads]{llncs}
\usepackage{graphicx}
\usepackage{cite}
\usepackage{listings}
\usepackage{color}
\usepackage{makeidx}  
\usepackage{graphicx}
\usepackage[footnote,draft,silent,nomargin]{fixme} 
\begin{document}
\title{An Analytics Tool for Exploring Scientific Software and Related Publications}
\titlerunning{SciSoftX: Exploring Scientific Software and Publications}
%
\author{Anett Hoppe\inst{1}\orcidID{0000-0002-1452-9509} \and
Jascha Hagen\inst{2} \and Helge Holzmann\inst{3}\orcidID{0000-0003-4811-6902} \and G\"unter Kniesel\inst{4} \and Ralph Ewerth\inst{1,3}\orcidID{0000-0003-0918-6297}}
%
%
\institute{German National Library of Science and Technology, Welfengarten 1B, 30167 Hannover, Germany, [firstname].[lastname]@tib.eu
\and
Leibniz Universit\"at Hannover, Germany, jascha\_hagen@yahoo.de
\and
L3S Research Center, Appelstrs. 9a, 30167 Hannover, Germany, holzmann@l3s.de
\and
University of Bonn, Bonn, Germany, gk@iai.uni-bonn.de
}

\maketitle              
\begin{abstract}
Scientific software is one of the key elements for reproducible research. However, classic publications and related scientific software are typically not (sufficiently) linked, and it lacks tools to jointly explore these artefacts. In this paper, we report on our work on developing an analytics tool for jointly exploring software and publications. The presented prototype, a concept for automatic code discovery, and two use cases demonstrate the feasibility and usefulness of the proposal. 
\end{abstract}

\section{Introduction}
\label{sec:intro}
The open science movement works towards the general availability of scientific insight and is considered one answer to the so-called "reproducibility crisis" \cite{baker2016}.  
Science results are often generated by a combination of software, data, and parameter settings, all of which contribute to the final result (and its interpretation). The complexity of all these elements is hardly describable in a single article -- and often the publication does not allow the full reproduction of the achieved results. 
In the line of work towards consequent reproducibility of scientific results, there are three main tasks to be tackled: 
    (a) motivate researchers to reproduce past results; 
    (b) develop novel ways for the integrated presentation of future scientific results;
    (c) develop tools which allow for easy exploration of existing scientific works.

The work at hand focusses on the two latter objectives. It presents a tool which facilitates the examination of existing research involving software by joint exploration of scientific article and respective source code. The prototype allows the exploration of both in one interface, and the semi-automatic creation of semantic relations between them. The software is extended by basic visualisations.

This kind of work is related to research areas, which have been active for decades: (a) automatic code analysis, and (b) the automatic analysis of scientific publications. Solutions for automatic code analysis aim at generating textual documentation (e.g. \cite{moser2016}), summarising code (e.g. \cite{nazar2016}), or at generating visualisations (e.g. \cite{chen2012}). Also common is the generation of formal code models using semantic technologies (e.g. \cite{atzeni2017}) or logical constructs 
, as realised in tools such as JTransformer\footnote{\url{http://sewiki.iai.uni-bonn.de/research/jtransformer/start}}.
 
While there is much work on linking code to other (textual) resources (e.g. traceability \cite{borg2014}), to documentation (e.g. \cite{chen2012}),  
on the automatic understanding of scientific publications (e.g. \cite{constantin2014}), or on linking publications with software and archiving them \cite{holzmann2016}, there has been little work on \textit{joint} analytics of scientific software and publications yet (e.g. \cite{witte2008}). 

\section{SciSoftX: Scientific Software Explorer}

The Software Explorer provides researchers with functionalities for the exploration of external article-software ensembles and/or annotation of own works for better comprehensibility. Its final version will provide functionalities such as (a) manual annotation of article-software relations; (b) semi-automatic discovery of relations; (c) visualisations for relation exploration.

\begin{figure}
\centering
\includegraphics[width=.9\textwidth]{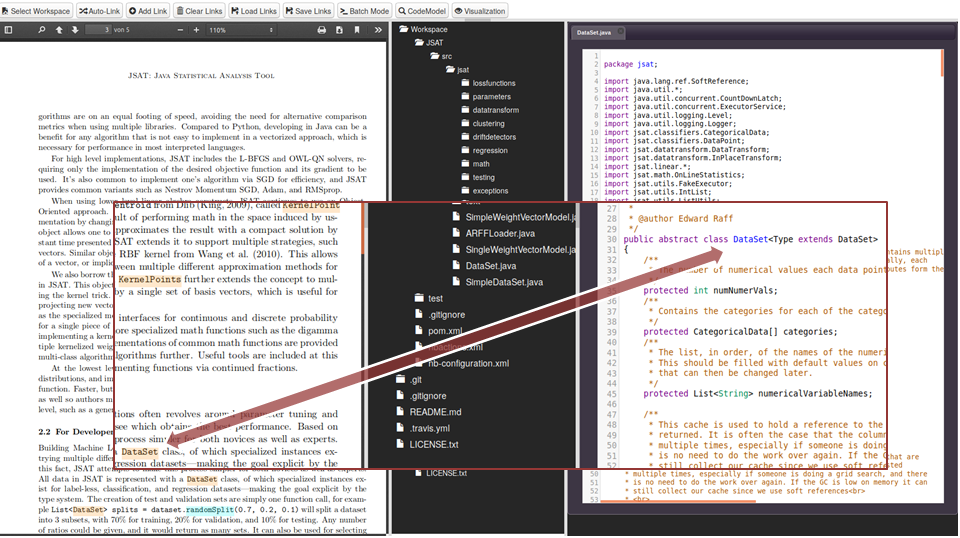}
\caption{Main window of the GUI: Linked code references are highlighted in colour.}
\label{fig:gui}
\end{figure}
\subsection{Functionality}
 SciSoftX allows the user to open and simultaneously view a software project and a publication (Fig. \ref{fig:gui}). Parsing and processing of source code is realised using ANTLR\footnote{\url{http://www.antlr.org/}} (Another tool for Language Recognition) that supports most of the relevant programming languages, while publications are processed via PDF.js\footnote{https://mozilla.github.io/pdf.js/}. The user can manually link code identifiers to relevant locations in the publication. When the user moves the mouse over a linked identifier in the publications, a tool tip shows the relevant source code positions. 

\textbf{Automatic Discovery of Code Identifiers and Snippets: } 
At the current stage, the tool contains a very basic method for code-relevant text snippets: It relies on the common convention of setting code elements in monospace fonts. The found identifiers are used to search the code model produced by ANTLR, multiple finds are disambiguated based on vicinity. In a random sample of 24 articles from computer science, the monospace-based linker was able to correctly detect 89,9\% of the links annotated by a human expert.

\textbf{Manual Annotation of Links: }
As a facilitator of exchange between scientists the tool also allows for the manual annotation of resources. In a step-wise process, the user marks article snippets, code elements and types the established link with one of the pre-defined labels. The created set of links can be exported to an xml format and imported by an interested reader.

 \begin{figure}[t]
  \centering
  \begin{minipage}{.42\columnwidth}
    \centering
      \includegraphics[width=\textwidth]{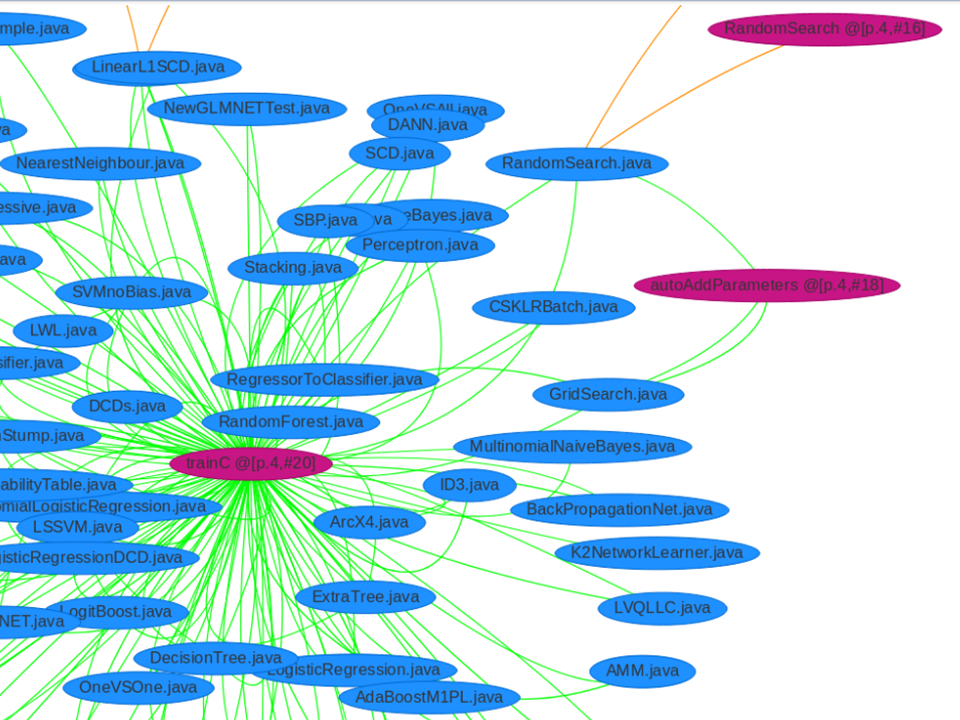}
  \end{minipage}
  \quad
  \begin{minipage}{.42\columnwidth}
    \centering
      \includegraphics[width=\textwidth]{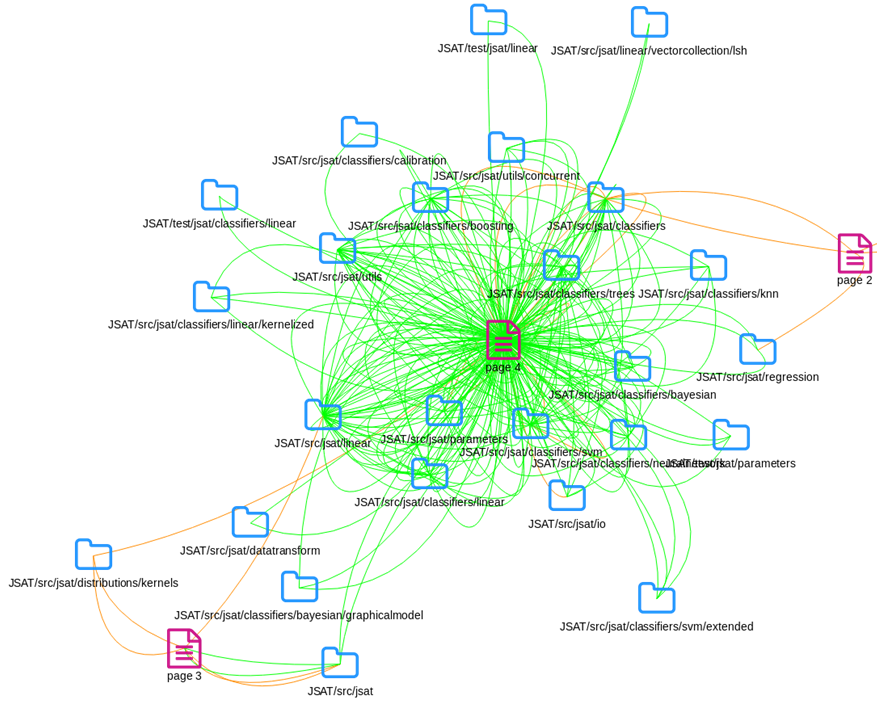}
  \end{minipage}

    \caption[]{Graph-based views on connections between software and publication. Red nodes: Mentions in publications. Blue nodes: Source code files (left) or packages (right).}
    \label{fig:vis}
\end{figure}
\textbf{Visualisations:} Graph-based visualisations illustrate relations between software and publication on different levels of abstraction. Fig. \ref{fig:vis} (left) depicts connections between source code and class files with exact positions in the publication (that is, page and line number); Fig. \ref{fig:vis} (right) displays connections at the package (software component) and page (publication) level.

\subsection{Use cases}
 
\textbf{Use case 1 -- Reader-side: } 
A researcher reads a publications that refers to a blob of software. She/he tries to understand the structure and rationale of the software. This time-consuming task can be supported by the automatic creation of links between textual description and actual source code, and the visualisations provided by SciSoftX. The user can click on nodes in the visualisation or on text elements that are highlighted in the publication and explore the implementation details, discover additional parameters, and understand the relevant code part step by step. Furthermore, she/he can manually add and save useful information and metadata, which can help future users to explore the software.

\textbf{Use case 2 -- Author-side: } 
A paper author wishes that his/her software can be easily understood, e.g. in a reviewing process or for re-use. Therefore, she/he uses the SciSoftX's manual and automatic methods to annotate the semantic relations between his/her paper and the underlying software and publishes the annotations. The visualisation of cross-modal relations can aid the author (and the reviewer) to decide whether all relevant code parts and parameters are covered by the publication. In this way, the tool helps to evaluate the quality of the software description in a paper.    
\section{Conclusion}
Reproducibility is one of the major issues of today's scientific landscape. In this paper, we have reported on work in progress for an analytics tool that allows users to explore relations between scientific software and publications. To this date, the tool features simple mechanisms for detecting links between software and publications who serve as a proof of concept. 

Future works will explore (a) more powerful infrastructures for code analysis 
; (b) more sophisticated means for text/image analysis, e.g. mapping diagrams and formulas to source code. 

\bibliographystyle{splncs04}
\bibliography{ecir2017}

\end{document}